\documentclass[a4paper,11pt]{article}
\usepackage{pos}
\usepackage{xcolor}

\def\bsp#1\esp{\begin{split}#1\end{split}}
\def\bal#1\eal{\begin{align}#1\end{align}}

\newcommand\rs   {\ensuremath{\mathrm{s}}}

\newcommand\rL   {\ensuremath{\mathrm{L}}}

\newcommand\GeV  {\ensuremath{\mathrm{GeV}}}

\newcommand\Mpl {\ensuremath{{M}_{\rm Pl}}}
\newcommand\mtp {\ensuremath{{M}_{\rm t}}}
\newcommand\mwp {\ensuremath{{M}_W}}
\newcommand\SARAH {{\tt SARAH}}
\newcommand\SPheno {{\tt SPheno}}
\newcommand\HiggsBounds {{\tt HiggsBounds}}

\title{Vacuum stability and scalar masses in the superweak extension of the standard model}
\ShortTitle{Scalar masses in SWSM}

\author[1]{Zoltán Péli}
\affiliation{Institute for Theoretical Physics, ELTE E\"otv\"os Lor\'and University,\\
P\'azm\'any P\'eter s\'et\'any 1/A, H-1117 Budapest, Hungary \\and\\
University of Debrecen,\\
P.O. Box 5, H-4010 Debrecen, Hungary}
\note{This research was supported by grant K-125105 of the National Research, Development and Innovation Fund in Hungary.}

\emailAdd{peli.zoltan@science.unideb.hu}

\abstract{
We summarize 
our analysis for the vacuum stability of the superweak extension of the standard model. 
The parameter space allowed by the $W$-boson mass measurements and collider searches for a new scalar particle are also presented.  
The final result is a well defined non-vanishing portion of the parameter space, where the vacuum is stable, the couplings are perturbative and is not excluded by $W$ mass measurements and collider searches.
}

\FullConference{
  41st International Conference on High Energy physics - ICHEP2022\\
  6-13 July, 2022\\
  Bologna, Italy
}

\begin{document}
\maketitle

\section{Introduction}
The superweak extension of the standard model (SWSM) has the potential to explain the origin of (i)
neutrino masses and mixing matrix elements \cite{Iwamoto:2021wko} and \cite{ichep2022:tk} in this conference, (ii) dark matter \cite{Iwamoto:2021fup} and \cite{ichep2022:sk} in this conference, (iii) cosmic inflation
\cite{Peli:2019vtp}, (iv) stabilization of the electroweak vacuum \cite{Peli:2019vtp} and possibly (v) leptogenesis (under investigation). This model is a beyond the standard model extension adding
one layer of interactions below the hierarchic layers of the strong, electromagnetic and weak forces, which is called superweak (SW) force \cite{Trocsanyi:2018bkm}, mediated by a new U(1) gauge boson $Z'$. 
In order to explain the origin of neutrino masses, the field content is enhanced by three generations of right-handed neutrinos 
(with Yukawa couplings $y_x$).
The new gauge symmetry is broken spontaneously by the vacuum expectation
value $w$ of a new complex scalar singlet.

We focus on (iv), improving significantly on our
previous work \cite{Peli:2019vtp} on constraining the parameters of the scalar sector in the model. Similar analyses 
have already been performed focusing on vacuum stability \cite{Falkowski:2015iwa, Robens:2015gla}.
In contrast to those studies, we use higher precision for computing the parameter space with stable vacuum and also investigate the effect of the sterile-neutrino Yukawa couplings.

We used the following high energy physics programs
(i) an SWSM model file  \cite{Iwamoto:2021wko} for \SARAH\ \cite{Staub:2013tta} to generate the two-loop renormalization group equations (RGE) and model files for \SPheno~\cite{Porod:2003um,Porod:2011nf}. Then (ii) \SPheno\ is utilized to compute two-loop Higgs boson masses. Finally, we generated (iii) input files for \HiggsBounds~\cite{Bechtle:2020pkv}
to obtain exclusion bands by direct Higgs searches.
\section{Vacuum stability}
\begin{figure}[t!]
\begin{center}
\includegraphics[width=0.40\linewidth]{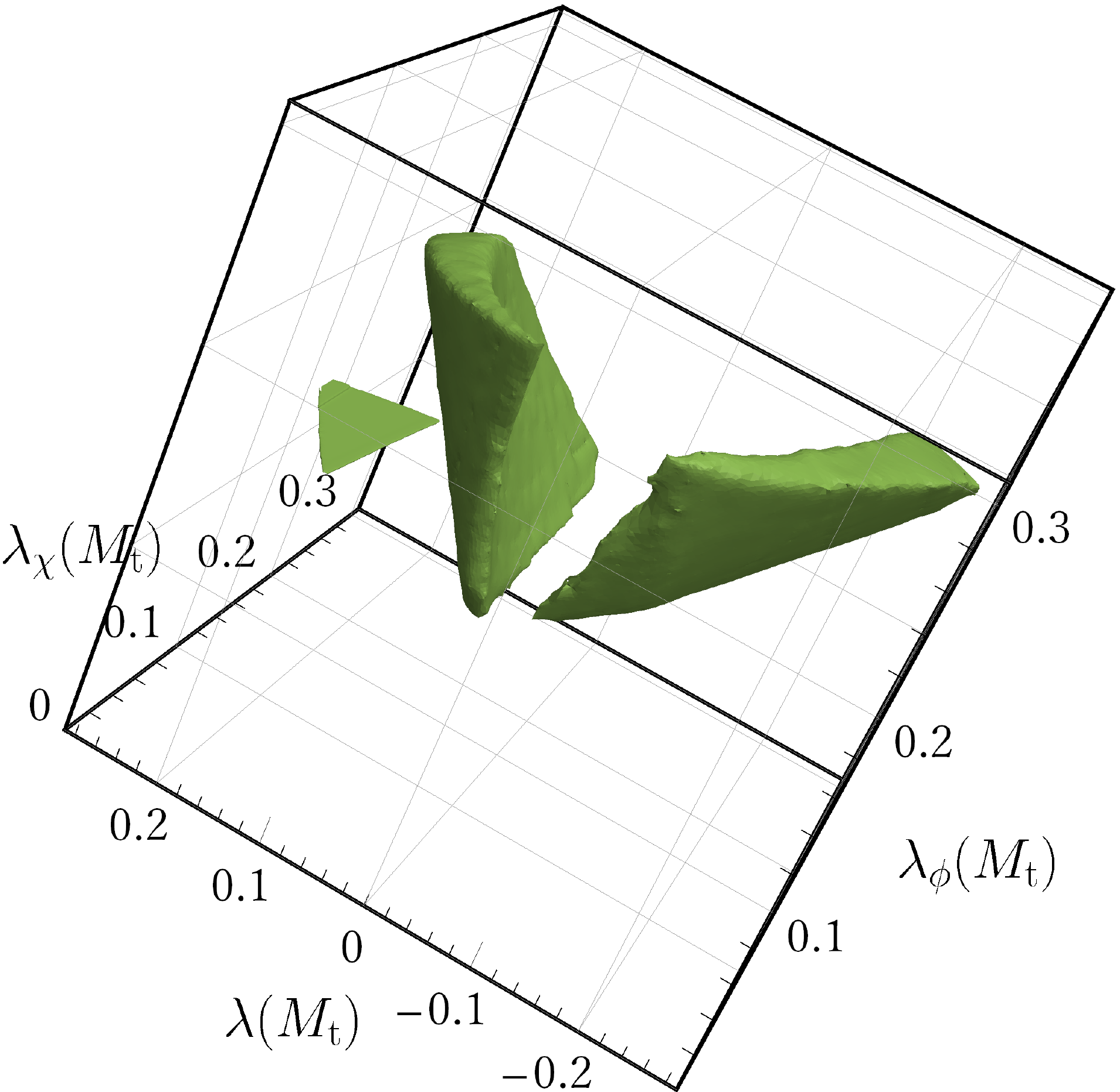}
\end{center}
\caption{\label{fig:0} 
3d slice of the parameter space allowed by vacuum stability, perturbativity and the existence of the new VEV $w$ computed at $y_x(\mtp)=0.4$.
}
\end{figure}

The scalar sector of the superweak extension is defined in \cite{Trocsanyi:2018bkm}. The scalar potential is 
\begin{equation}
\label{eq:V_phichi}
    V(H,S) = V_0 - \mu_\phi^2 H^2 - \mu_\chi^2 S^2
    +\lambda_\phi H^4 + \lambda_\chi S^4
    +\lambda H^2 S^2
    \,,
\end{equation}
where the fields $H$ and $S$ mix into the mass eigenstates $h$ and $s$ with the mixing angle $\theta_\rs$.
We check the vacuum stability of the potential \eqref{eq:V_phichi} and the perturbativity of the couplings from the scale of the top quark mass $\mtp$ up to the Planck mass, i.e. $\mu\in (\mtp, \Mpl)$ by solving the two-loop renormalization group equations. We scan the four dimensional parameter space $ \lambda_\phi(\mtp),\lambda_\chi(\mtp),\lambda(\mtp), y_x(\mtp)$ and extract the corresponding $w(\mtp)$ at two-loop accuracy. We only accept parameter points if $w(\mtp) > 0$ in order to obtain non-trivial phenomenology. A three dimensional slice of the allowed parameter space at $y_x(\mtp)=0.4$ is shown in Fig.~\ref{fig:0}.

\section{Predictions for the W mass in the SWSM}
In the $\overline{\text{MS}}$ 
scheme the scalar SW contribution to the $W$-boson mass $M_W$ is the sum of the coupling shift effects $\delta g_\rL, \delta v$ and the SWSM self-energy correction $\Pi^{\text{SW}}_{WW}$ given by 
\begin{equation}\label{app:1loopwmass}
    \mwp = \mwp^{\text{theo.}} + \delta M_W,
    \quad\text{with}\quad
      \delta M_W = \mwp \frac{\delta v}{v} + \frac{1}{2}\delta g_\rL v + \frac{1}{2}\frac{\Pi^{\text{SW}}_{WW}(\mwp)}{\mwp}, 
\end{equation}
where we use $\mwp^{\text{SM. theo.}} = 80.360 \pm 0.012 ~\GeV$ as the theoretical prediction for W-boson mass in the SM. The world average of the experiments
\cite{ParticleDataGroup:2022pth} is $M_W^{\text{exp.}} = (80.377 \pm 0.012)\,\GeV\,$, ignoring the recent result from CDFII \cite{CDF:2022hxs}.
The SWSM contribution $\delta M_W$ is negative for $M_s > M_h$ and positive for  $M_s < M_h$.

\begin{figure}[t!]
\begin{center}
{
\includegraphics[width=0.40\linewidth]{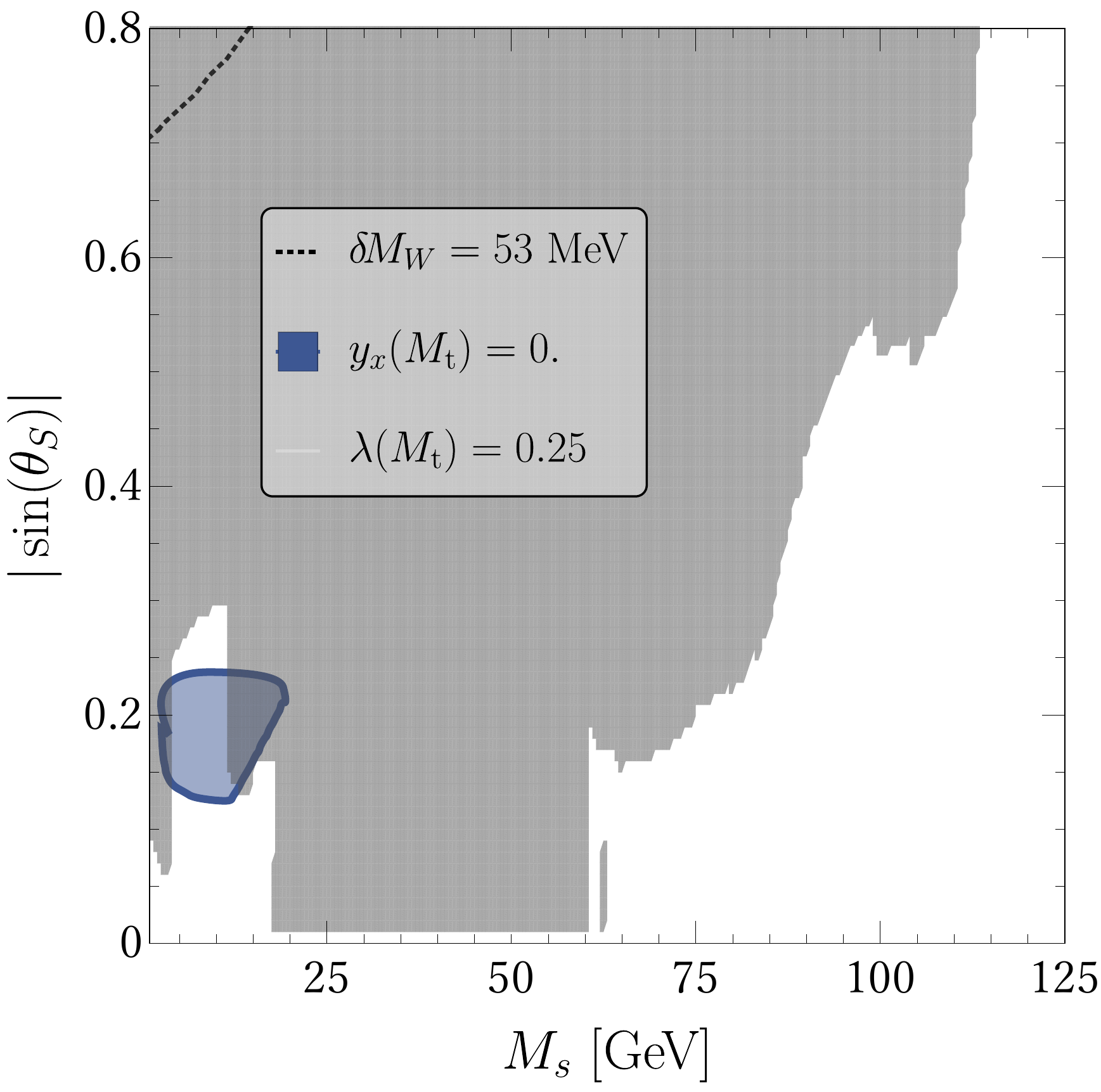}
\includegraphics[width=0.40\linewidth]{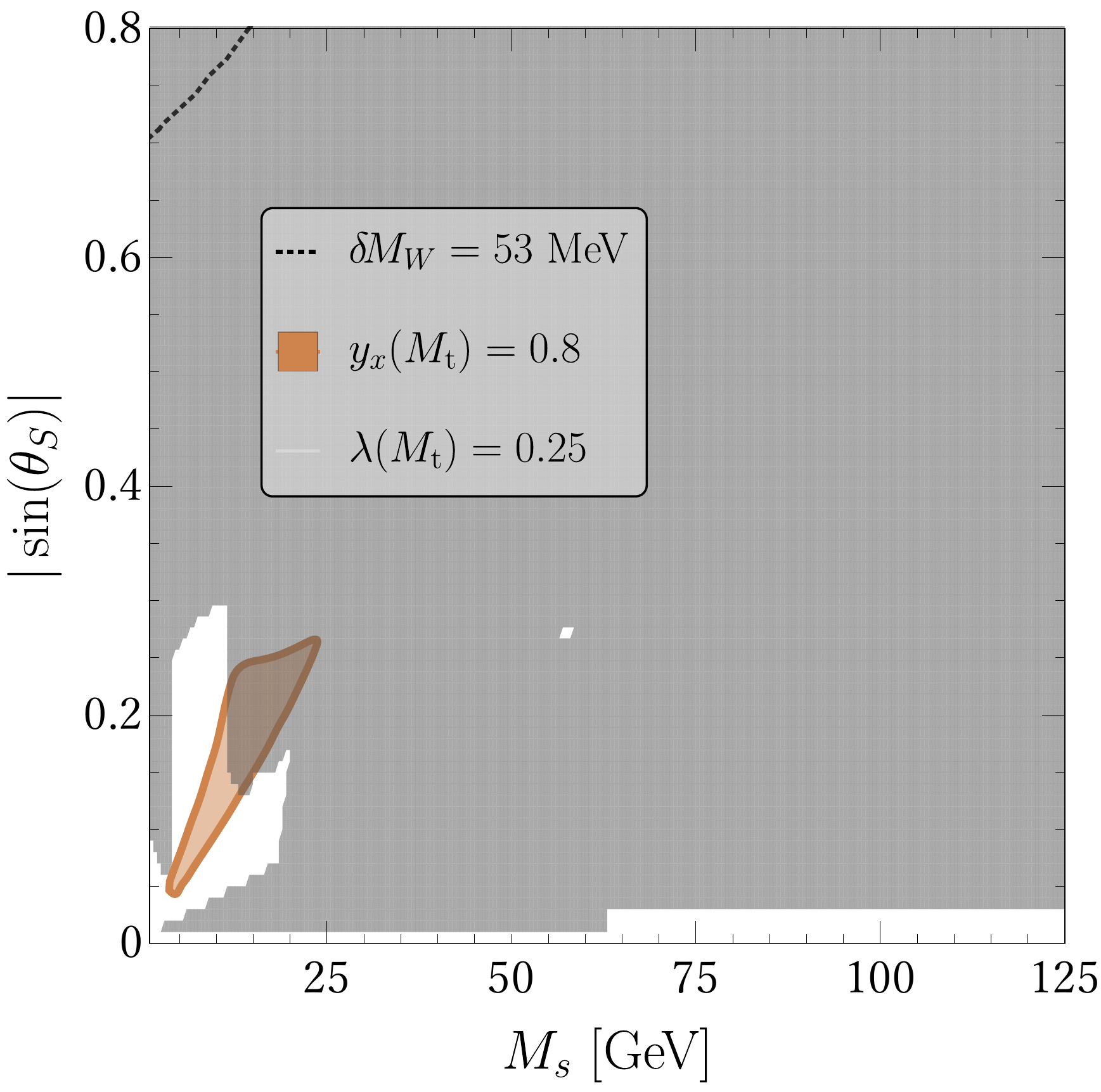}
}
\end{center}
\caption{\label{fig:1} 
Colored regions: allowed by vacuum stability and perturbativity at different $y_x(\mtp)$. Above the dashed line: excluded by $W$-boson mass measurements. Gray region: excluded by collider searches (\HiggsBounds~5).
}
\end{figure}
\section{Results and outlook}
We found a well-defined, allowed parameter space for both $M_s > M_h$ (see \cite{ichep2022:tz} in this conference) and $M_s < M_h$ (see Fig.~\ref{fig:1}). The left figure corresponds to decoupled sterile neutrinos, while the right figure corresponds to a sizeable sterile neutrino Yukawa coupling, where  
$y_x(\mtp)\gtrsim 0.9$ is excluded by the vacuum stability and perturbativity analysis. 
The colored region of the parameter space corresponds to the perturbativity of the couplings and stable vacuum, while the gray region is excluded by direct Higgs searches, computed by \HiggsBounds~5. At this point, our findings suggest, that experimental probes for a scalar particle with higher sensitivity in the mass range $1~\GeV - 20~\GeV$ is well motivated. However, these predictions 
have to be scrutinized severely, i.e.~confronted also with coupling measurements of the $125~\GeV$ Higgs boson.
%


\begin{thebibliography}{99}
\bibitem{Iwamoto:2021wko}
S.~Iwamoto, T.~J.~K\"arkk\"ainen, Z.~P\'eli and Z.~Tr\'ocs\'anyi,
{\it One-loop corrections to light neutrino masses in gauged $U(1)$ extensions of the standard model},
Phys. Rev. D \textbf{104} (2021)055042 
[2104.14571].
%
\bibitem{ichep2022:tk}
T.~J.~K\"arkk\"ainen,
{\it Neutrino physics from a gauged $U(1)$ extension of the Standard Model}, contribution to the 41st International Conference on High Energy physics, in this proceedings. 
%
\bibitem{Iwamoto:2021fup}
S.~Iwamoto, K.~Seller and Z.~Tr\'ocs\'anyi,
{\it Sterile neutrino dark matter in a U(1) extension of the standard model},
JCAP \textbf{01} (2022) 035
[2104.11248].
%
\bibitem{ichep2022:sk}
K.~Seller,
{\it Sterile neutrino dark matter in the super-weak model},
contribution to the 41st International Conference on High Energy physics, in this proceedings [arXiv:2210.16090 [hep-ph]].
%
\bibitem{Peli:2019vtp}
Z.~P\'eli, I.~N\'andori and Z.~Tr\'ocs\'anyi,
{\it Particle physics model of curvaton inflation in a stable universe},
Phys. Rev. D \textbf{101} (2020) 063533
[1911.07082].
%
\bibitem{Trocsanyi:2018bkm}
Z.~Tr\'ocs\'anyi,
{\it Super-weak force and neutrino masses},
Symmetry \textbf{12} (2020) 107
[1812.11189].
%
\bibitem{Falkowski:2015iwa}
A.~Falkowski, C.~Gross and O.~Lebedev,
{\it A second Higgs from the Higgs portal},
JHEP \textbf{05} (2015), 057
[1502.01361].
%
\bibitem{Robens:2015gla}
T.~Robens and T.~Stefaniak,
{\it Status of the Higgs Singlet Extension of the Standard Model after LHC Run 1},
Eur. Phys. J. C \textbf{75} (2015), 104
[1501.02234].
%
\bibitem{Staub:2013tta}
F.~Staub,
{\it SARAH 4 : A tool for (not only SUSY) model builders},
Comput. Phys. Commun. \textbf{185} (2014), 1773-1790
[1309.7223].
%
\bibitem{Porod:2003um}
W.~Porod,
{\it SPheno, a program for calculating supersymmetric spectra, SUSY particle decays and SUSY particle production at e+ e- colliders},
Comput. Phys. Commun. \textbf{153} (2003), 275-315
[hep-ph/0301101].
%
\bibitem{Porod:2011nf}
W.~Porod and F.~Staub,
{\it SPheno 3.1: Extensions including flavour, CP-phases and models beyond the MSSM},
Comput. Phys. Commun. \textbf{183} (2012), 2458-2469
[1104.1573].
%
\bibitem{Bechtle:2020pkv}
P.~Bechtle, D.~Dercks, S.~Heinemeyer, T.~Klingl, T.~Stefaniak, G.~Weiglein and J.~Wittbrodt,
{\it HiggsBounds-5: Testing Higgs Sectors in the LHC 13 TeV Era},
Eur. Phys. J. C \textbf{80} (2020) 1211
[2006.06007].
%
\bibitem{ParticleDataGroup:2022pth}
R.~L.~Workman \textit{et al.} [Particle Data Group],
{\it Review of Particle Physics},
PTEP \textbf{2022} (2022), 083C01
%
\bibitem{CDF:2022hxs}
T.~Aaltonen \textit{et al.} [CDF],
{\it High-precision measurement of the W boson mass with the CDF II detector},
Science \textbf{376} (2022) 170-176
%
\bibitem{ichep2022:tz}
Z.~Tr\'ocs\'anyi,
{\it SWSM phenomenolgy},
contribution to the 41st International Conference on High Energy physics, in this proceedings. 
\end{thebibliography}

\end{document}